\documentclass[journal]{IEEEtran}
\usepackage{cite,amsmath,algorithmic,booktabs,graphicx,subfigure,xcolor,multirow,url,amssymb}
\usepackage[lined,ruled,commentsnumbered]{algorithm2e}

\begin{document}

\title{Front-end Replication Dynamic Window (FRDW) \\ for Online Motor Imagery Classification}

\author{Xinru~Chen, Jiayu~An, Huanyu~Wu, Siyang~Li, Bin~Liu, and Dongrui~Wu
\thanks{X.~Chen, J.~An, H.~Wu, S.~Li and D.~Wu are with the Ministry of Education Key Laboratory of Image Processing and Intelligent Control, School of Artificial Intelligence and Automation, Huazhong University of Science and Technology, Wuhan 430074, China. They are also with the Shenzhen Huazhong University of Science and Technology Research Institute, Shenzhen, China.}
\thanks{B.~Liu is with Beijing Fenghuo Wanjia Technology Co. Ltd., Beijing, 100026, China.}
\thanks{X.~Chen and J.~An contributed equally to this work.}
\thanks{D.~Wu is the corresponding author (e-mail: drwu09@gmail.com).}}

\maketitle

\begin{abstract}
Motor imagery (MI) is a classical paradigm in electroencephalogram (EEG) based brain-computer interfaces (BCIs). Online accurate and fast decoding is very important to its successful applications. This paper proposes a simple yet effective front-end replication dynamic window (FRDW) algorithm for this purpose. Dynamic windows enable the classification based on a test EEG trial shorter than those used in training, improving the decision speed; front-end replication fills a short test EEG trial to the length used in training, improving the classification accuracy.  Within-subject and cross-subject online MI classification experiments on three public datasets, with three different classifiers and three different data augmentation approaches, demonstrated that FRDW can significantly increase the information transfer rate in MI decoding. Additionally, FR can also be used in training data augmentation. FRDW helped win national champion of the China BCI Competition in 2022.
\end{abstract}


\section{Introduction}

A brain-computer interface (BCI) can measure and process the subject's brain activities and translate them into interactive information or commands for external device control \cite{Lotte2018}. Sensorimotor rhythm (SMR) based BCIs \cite{Andrea2005,Vaughan2006,Yuan2014} are based on the principle that the execution or imagination of limb movements, the latter also known as motor imagery (MI), changes the cortical rhythmic activity \cite{Steriade2005}. The increase and decrease of the SMR are called event-related synchronization (ERS) and event-related desynchronization (ERD), respectively.

Electroencephalogram (EEG) can record SMR changes during MI, which can be used for BCI control \cite{Pfurtscheller2006a,Pfurtscheller2006b,Neuper2006}. MI-based BCIs are achieved by imagining the movement of a specific body part, e.g., left hand, right hand, feet, or tongue, without actually performing it \cite{Lotze2006}. They have been used in  wheelchair/robot control, stroke rehabilitation, gaming, etc \cite{Coyle2011,LaFleur2013,Fernndez2016,Ron-Angevin2017,Lpez-Larraz2018}.

MI classification could be performed offline or online. Offline classification means the entire test EEG data are available offline for analysis. Online classification collects the subject's EEG signals and makes inferences in real-time, aiming at both high classification accuracy and fast response. The information transfer rate (ITR) is an important metric for online BCIs.

There are several challenges for online MI classification:
\begin{enumerate}
\item\emph{Varying EEG trial length}. Most EEG classification algorithms assume the input trial has a fixed length, whereas for fast response, EEG trials in online classification usually have varying lengths.
\item\emph{Trade-off between speed and accuracy}. For fast response, classifications should be made for short EEG trials, which usually reduces the accuracy.
\item\emph{Large individual differences}. In cross-subject MI classification, to alleviate individual differences, some EEG data from the test subject are usually needed for model calibration. Online classification usually has less calibration data than offline classification.
\end{enumerate}

To cope with these challenges, we propose a front-end replication dynamic window (FRDW) approach for online MI classification. Our main contributions are:
\begin{enumerate}
\item We propose a front-end replication (FR) approach to fill each test EEG trial to a fixed length required by the online classifier, improving the classification accuracy. Additionally, FR can also be used in training data augmentation.
\item We use a dynamic window (DW) approach to adaptively adjust the length of each test EEG trial to improve the decoding speed.
\item We integrate Euclidean alignment (EA) \cite{he2019} with FRDW to accommodate individual differences in online cross-subject MI classification.
\end{enumerate}
Extensive within- and cross-subject experiments on three public MI datasets with three classifiers and three data augmentation approaches demonstrated the effectiveness of our proposed FRDW approach. FRDW helped win national champion of the China BCI Competition in 2022\footnote{http://www.worldrobotconference.com/cn/view/1938.html}.

The remainder of this paper is organized as follows: Section~II introduces related works. Section III describes the FRDW approach. Section IV introduces the experimental settings. Section V presents the experimental results. Finally, Section VI draws conclusions.

\section{Related Works}

This section introduces related works on offline MI classification, online MI classification, and DW approaches for EEG classification.

\subsection{Offline MI Classification}

Both conventional machine learning and deep learning (DL) have been used in offline EEG-based MI classification.

Conventional machine learning typically uses expert knowledge to extract EEG features and then feeds them into a traditional classifier, e.g., support vector machine (SVM). Common spatial pattern (CSP) \cite{Lotte2010} and its variants, e.g., sparse CSP \cite{Arvaneh2011}, L1-norm-based CSP \cite{Wang2011}, divergence CSP \cite{Samek2013}, and probabilistic CSP \cite{Wu2014}, have been widely used in MI signal processing and feature extraction. Filter bank CSP \cite{ang2008} first partitions the EEG signal into different frequency bands, and then applies CSP to each of them. Some studies also used power spectral density \cite{Kim2018, Alam2021} or wavelet transform \cite{Xu2018} features.

Most DL approaches take raw EEG signals as the input. Popular convolutional neural network (CNN) based DL approaches include EEGNet \cite{lawhern2018}, Deep ConvNet \cite{schirrmeister2017} and Shallow ConvNet \cite{schirrmeister2017}. EEG-TCNet \cite{ingolfsson2020}, TCNet-Fusion \cite{Musallam2021}, and ATCNet \cite{altaheri2022} are their enhancements by temporal convolution, fusion layer, and attention. Filter bank multi-scale CNN (FBMSNet) \cite{Liu2023} extends filter bank from traditional learning to DL.

\subsection{Online MI Classification}

Online MI classification has been attracting much attention, due to the requirement of real-time BCIs.

Yang \textit{et al.} \cite{fengwei2023} applied data augmentation to the first three-second of a test trial to get five samples, and then voted the five real-time predictions for the final result. The average binary classification accuracy on 80 test trials from two subjects was $71.3\%$.  Tayeb \textit{et al.} \cite{tayeb2019} controlled the movement of a robotic arm in real-time, which only responded to high-confidence predictions. Parashiva \textit{et al.} \cite{parashiva2022} trained an Error-Related Potential detection model to learn the brain response to feedback and make automatic corrections, achieving $64.88\%$ online classification accuracy. Furthermore, asynchronous MI \cite{kus2012,choi2020} first recognizes whether the subject is performing MI, and then makes classification. Its typical performance measure includes both classification accuracy and ITR.

\subsection{Dynamic Window for EEG Classification}

DWs were introduced to alleviate the limitations of fixed windows (FWs). They have found applications in online Steady-State Visual Evoked Potential (SSVEP) based BCIs.

Spatiotemporal equalization DW recognition \cite{yang2018} allows the adaptive control of the stimulus timing while maintaining high recognition accuracy, significantly improving the ITR and the system's adaptability to different subjects. Chen \textit{et al.} \cite{chen2021} proposed a filter bank canonical correlation analysis based training-free DW recognition approach for SSVEP. Hadi \textit{et al.} \cite{Habibzadeh2021} proposed a novel DW classifier, using ensembling learning for SSVEP recognition. \cite{zhou2022a,Yin2022} enhanced DW threshold selection to further improve the ITR for SSVEP classification. DL-based DW has also been used in SSVEP recognition, e.g., Zhou \textit{et al.} \cite{Zhou2022b} proposed an EEGNet-DW approach, which uses a different EEGNet model and threshold for each DW length.

To our knowledge, there has not been DW approaches for MI-based BCIs.

\section{Methodology}

This section introduces our proposed FRDW algorithm. The Python code is available at https://github.com/XinRu2001/FRDW.

\subsection{Problem Setting}

Assume we have trained an MI classifier $f(X)$, where $X\in \mathbb{R}^{C\times N}$, in which $C$ is the number of EEG channels, and $N$ the fixed trial length. During the test stage, $n$ test trials $\{X'_{i}\}_{i=1}^{n}$ arrive one by one. For each test trial, each sampling update brings in $L'$ points, i.e., each test trial arrives in the sequence of dimensionality $\mathbb{R}^{C \times L^\prime}$, $\mathbb{R}^{C \times 2L^\prime}$, $\ldots$, $\mathbb{R}^{C \times N}$. For maximum ITR, we should make an accurate classification as early as possible, not necessarily waiting for the full $X_i'\in\mathbb{R}^{C \times N}$.

\subsection{FRDW for Online Within-Subject MI Classification}

We propose FRDW to improve the ITR of MI-based BCIs.

A DW on the test dataflow acts on the currently available test trial data ${X' \in \mathbb{R}^{C \times L}}$, where $\underline{L}\le L\le N$ and $\underline{L}$ is the minimum length. When $L<N$, FR is used to fill $X'$ to length $N$ to match the trial length used in training. Specifically, ${X'}$ is repeatedly concatenated:
\begin{align}
\overline{X} = [{X'}, {X'}, \cdots, {X'}].   \label{eqFR}
\end{align}
$\overline{X}$ is then trimmed to have $N$ columns and used as the input to the classifier. This approach is called FR because trimming keeps the front-end of $X'$.

When the maximum classification probability on the current $\overline{X}$ is lower than a confidence threshold $\tau$, the current window may be too short for reliable classification. In such case, we wait for another ${L^\prime}$ samples, and apply FR to the test input of length $L+{L}'$. The process is repeated until the maximum classification probability exceeds or equals $\tau$. In the worst-case scenario, the entire test trial of length $N$ is used for classification.

Algorithm~\ref{Alg:FRDW} gives the pseudo-code of FRDW for online within-subject MI classification.

\begin{algorithm}[h]
    \caption{FRDW for online within-subject MI classification.}     \label{Alg:FRDW}
    \KwIn{${X' \in \mathbb{R}^{C \times {L}}}$, the current test trial;\\
    \hspace*{0.95cm} $\underline{L}$, the minimum trial length;\\
    \hspace*{0.95cm} $(p,m)=f(X)$, the classifier, where $X\in \mathbb{R}^{C\times N}$, $p$ is the maximum prediction probability, and $m$ is the predicted class;\\
    \hspace*{1cm} $\tau$, the confidence threshold.}
    \KwOut{$m$, the predicted class.}
    \While{$L < N+{L}'$}{
        \uIf{$L<\underline{L}$}
            {$L\gets L+{L}'$;\\
            Get ${X' \in \mathbb{R}^{C \times L}}$ from the dataflow;\\}
            \uElseIf{$\underline{L} \leq L<N$}{
                Construct $\overline{X}$ using (\ref{eqFR});\\
                $(p,m) \gets f({\overline{X}})$;\\
                \eIf{$p\geq \tau$}{
                    \Return{$m$}.}
                    {$L\gets L+{L}'$;\\
                    Get ${X' \in \mathbb{R}^{C \times L}}$ from the dataflow;\\
                   }}
            \Else{
                $(p,m) \gets f(X_N)$, where $X_N$ contains the first $N$ points of $X'$;\\
                \Return{$m$}.}
        }
\end{algorithm}

\subsection{FRDW with EA for Online Cross-Subject MI Classification} \label{FRDW-EA}

In cross-subject MI classification, FRDW can be combined with EA \cite{he2019} for better performance. EA aligns EEG trials from different subjects in the Euclidean space to reduce individual differences, improving the transfer learning performance on a new subject \cite{drwuMITLBCI2022}.

Assume there are $n$ trials $\{X_i\}_{i=1}^n$ from a subject. EA first computes the reference matrix $\overline{R}$ as the arithmetic mean of all $n$ covariance matrices:
\begin{align}
\overline{R}=\frac{1}{n}\sum_{i=1}^{n}{{X_i}{{X}^{\top}_i}}.     \label{EA-R}
\end{align}
Each trial $X_i$ is then aligned by
\begin{align}
\widetilde{X}_i={\overline{R}^{-1/2}}{X_i}, \quad i=1,...,n    \label{EA-X}
\end{align}
where $\widetilde{X}_i$ is the aligned sample for $X_i$.

After EA, the mean covariance matrix of each subject equals the identity matrix $I$:
\begin{align}
	\frac{1}{n}\sum_{i= 1}^{n}\widetilde{X}_{i}\widetilde{X}_{i}^{T}&=\overline{R}^{-1/2}\left (\frac{1}{n}\sum_{i= 1}^{n}X_{i}X_{i}^{T} \right )\overline{R}^{-1/2}\nonumber\\
	&=\overline{R}^{-1/2}~\overline{R}~\overline{R}^{-1/2}= I. \label{EA-align}
\end{align}
i.e., EEG trials from different subjects become more consistent.

Since EA requires a few EEG trials for calculating the reference matrix $\overline{R}$, it raises problems at the beginning of the test phase when there are too few test trials. When the number of test trials is smaller than a threshold $n_{EA}$ ($n_{EA}=10$ in this paper), no EA is performed on the test trials, and a classifier without EA is applied to them. When $n_{EA}$ test trials are accumulated, we compute and save the reference matrix $\overline{R}$, perform EA on each test trial thereafter, and apply a classifier with EA to it.

Algorithm~\ref{Alg:FRDW with EA} shows the pseudo-code of FRDW with EA for online cross-subject MI classification.

\begin{algorithm}[h]
    \caption{FRDW with EA for online cross-subject MI classification.}\label{Alg:FRDW with EA}
    \KwIn{$X'_n \in \mathbb{R}^{C \times L}$, the current test trial;\\
    \hspace*{0.95cm} $\underline{L}$, the minimum trial length;\\
    \hspace*{0.95cm} $(p,m)=f(X)$, the classifier without EA, where $X\in \mathbb{R}^{C\times N}$, $p$ is the maximum prediction probability, and $m$ is the predicted class;\\
    \hspace*{0.95cm} $(p,m)=f_{EA}(X)$, the classifier with EA;\\
    \hspace*{0.95cm} $\tau$, the confidence threshold;\\
    \hspace*{0.95cm} $n_{EA}$, the minimum number of test trials for applying EA;\\
    \hspace*{0.95cm}$\{X'_{i}\}_{i=1}^{n-1}$, all test trials before the current trial;\\
    \hspace*{0.95cm} $\overline{R}$, the reference matrix of EA; needed only when $n>n_{EA}$.}
    \KwOut{$m$, the predicted class.}
        \If{$n \leq n_{EA}$}
        {
            $(p,m) \gets f ({X}_n')$, where ${X}_n'\in\mathbb{R}^{C\times N}$;\\
       }
        \If{$n = n_{EA}$}
        {
            Compute $\overline{R}$ on $\{X'_{i}\}_{i=1}^n$ using (\ref{EA-R});\\
       }
        \If{$n \ge n_{EA}$}
        {
            $\widetilde{X}_n'\gets \overline{R}^{-1/2}X_n'$;\\
                \While{$L < N+{L}'$}{
        \uIf{$L<\underline{L}$}
            {$L\gets L+{L}'$;\\
            Get ${X_n' \in \mathbb{R}^{C \times L}}$ from the dataflow;\\}
            \uElseIf{$\underline{L} \leq L<N$}{
                Construct $\overline{X}_n$ from $\widetilde{X}_n'$ using (\ref{eqFR});\\
                $(p,m) \gets f_{EA}({\overline{X}_n})$;\\
                \eIf{$p\geq \tau$}{
                    \Return{$m$}.}
                    {$L\gets L+{L}'$;\\
                    Get ${X_n' \in \mathbb{R}^{C \times L}}$ from the dataflow;\\
                   }}
            \Else{
                $(p,m) \gets f_{EA}(\widetilde{X}_N)$, where $\widetilde{X}_N$ contains the first $N$ points of $\widetilde{X}_n'$;\\
                \Return{$m$}.}        }       }
\end{algorithm}

\section{Experiments}

Extensive experiments were performed to validate the superior performance of FRDW.

\subsection{Dataset and Preprocessing}

Three public MI datasets were used in our experiments, whose statistics are summarized in Table~\ref{tab:datasets}.

\begin{table}[h]\centering \setlength\tabcolsep{1mm}
\caption{statistics of the three MI datasets.} \label{tab:datasets}
\begin{tabular}{@{}lcccc c@{}}
\toprule
Dataset & \# Subjects & \# Channels & \# Training Trials & \# Test Trials & \# Classes \\ \midrule
MI1  & 9          & 22         & 288            & 288           & 4         \\
MI2  & 9          & 22         & 144            & 144           & 2         \\
MI3    & 9          & 3          & 400            & 320           & 2         \\ \bottomrule
\end{tabular}
\end{table}

All three datasets were from BCI Competition \uppercase\expandafter{\romannumeral4} \cite{tangermann2012} and had the same collection protocol. Each subject sat in front of a computer screen. At the beginning of each trial, a fixation cross appeared on the screen, accompanied by a warning tone. Shortly after, an arrow pointing to a particular direction appeared as a cue (e.g., left arrow for left hand, down arrow for feet), prompting the subject to perform the instructed MI task until the fixation cross disappeared from the screen. The next trial started after a short break. EEG signals were recorded during the experiment.

EEG signals in all three datasets were imported from an open-source repository\footnote{http://www.bnci-horizon-2020.eu/database/data-sets}. Their specific properties are:
\begin{enumerate}
\item \emph{MI1}. The 22-channel (Fz, FC3, FC1, FCz, FC2, FC4, C5, C3, C1, Cz, C2, C4, C6, CP3, CP1, CPz, CP2, CP4, P1, Pz, P2, POz) EEG signals were sampled at 250 Hz from 9 subjects. Two sessions on different days were recorded for each subject, one for training and the other for test. Each training or test session contained 72 trials per class, and there were four classes (left hand, right hand, feet, and tongue).
\item \emph{MI2}. MI2 is identical to MI1, except that it only considered binary classification (left hand and right hand).
\item \emph{MI3}. The 3-channel (C3, Cz, C4) EEG signals were sampled at 250 Hz from 9 subjects for binary classification (left hand and right hand). There were three training sessions and two test sessions. The first two training sessions had 60 trials per class without feedback, and the last three sessions had 80 trials per class with smiley feedback.
\end{enumerate}

We detrended the raw full-channel EEG data and used a 5th-order 8-26Hz Butterworth bandpass filter to remove muscle artifacts and direct current drift.

\subsection{Data Augmentation}

We used two EEG data augmentation approaches to reduce model overfitting:
\begin{enumerate}
\item  \emph{overlap}. The trials are augmented using sliding windows, i.e., each sliding window contains 25 sampling points of the original data, and there is a 75-point overlap between two successive windows. Note that `\emph{none}' in our comparison means no overlap between any successive windows was used.
\item  \emph{FR}. FR introduced in Section III can also be employed for data augmentation in training. It uses sliding windows of length $0.7*N$ and 25-point overlap to segment each training trial. Then, for each sliding window, FR is used to complement the front-end data to the back, making the total length $N$.
\end{enumerate}

\subsection{Classifiers}

Three classifiers were considered:
\begin{enumerate}
\item \emph{EEGNet}. EEGNet \cite{lawhern2018} is a popular CNN for EEG classification. It starts with a temporal convolution followed by a depthwise convolution and a separable convolution, and finally a fully connected layer for classification. We used eight temporal filters and two spatial filters, and dropout rate $0.25$.

\item \emph{CSP+Transformer}. CSP \cite{Lotte2010}+Transformer \cite{vaswani2017} was inspired by Conformer \cite{song2022}, which used a convolutional transformer. However, the data length used by Conformer was about eight times longer than ours, and it is difficult for us to extract features with only two one-dimensional convolutions. Therefore, we utilized CSP to extract log-variance features and performed a temporal convolution and a depthwise convolution on the features as patch embeddings. The transformer encoder block was repeated three times, and a fully connected layer was used for classification. For 4-class MI classification, we used a one-versus-rest strategy for CSP \cite{song2021}, which divided the 4-class classification task into four binary-classification tasks. We concatenated the first four rows of each filter as the final filter.

\item \emph{CSP+SVM}. Feature extraction using CSP and then classification using SVM \cite{Johan1999} is a classical MI classification approach. We employed radial basis function kernel with regularization parameter $C = 0.1$ for within-subject SVM training, and linear kernel with $C = 1$ for cross-subject SVM training (default values were used for all other parameters).
\end{enumerate}

\subsection{Performance Metric}

The ITR, which considers both the classification speed and accuracy, was used as the primary performance matric:
\begin{align}
\mathrm{ITR}=\frac{60}{T}(\log_2M + P\log_2P+ (1-P)\log_2\frac{1-P}{M-1}),\label{eqitr}
\end{align}
where $T$ is the average trial length (s), $M$ the number of classes, and $P$ the classification  accuracy. The unit of ITR is bits/min. When $P<\frac{1}{M}$, i.e., the classification accuracy is lower than random, the ITR is set to 0.

\begin{figure}[h]    \centering
    \includegraphics[width=.9\linewidth,clip]{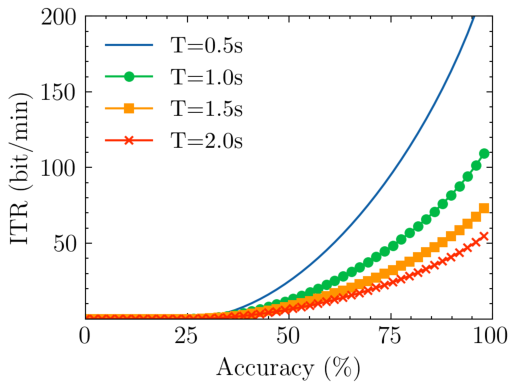}
    \caption{Relationship between the ITR and classification accuracy for different trial length $T$.}
    \label{itr}
\end{figure}

Taking four-class classification as an example, the relationship between the ITR and the classification accuracy for different $T$ is shown in Fig.~\ref{itr}. For a fixed $T$, the ITR grows exponentially as the accuracy increases. For a fixed accuracy, the ITR is an inverse function of the  trial length. So, to improve the ITR, we should increase the classification accuracy while reducing the trial length.

\subsection{Experimental Settings}

All datasets were partitioned into two parts, training and test. We reserved part of the training data as the validation set, i.e., each class's last 12 trials (the last block) on MI1 and MI2, and the last 44 trials ($20\%$) on MI3.

For within-subject MI classification, we first recorded the hyper-parameters (e.g., trial length, number of epochs) corresponding to the best ITR on the validation set, and then combined the training and validation sets to train the final model using them. For cross-subject MI classification, we used leave-one-subject-out cross-validation, i.e., the test set of one subject was used for testing and the training sets of all other subjects were combined for training. The model with the best validation ITR was used (no re-training as in the within-subject case).

During offline training, the same data preprocessing and augmentation procedures were applied to the training and validation sets. We used the training set to calculate the CSPs, which were then used for validation set feature extraction.

We also used offline data to simulate the online data acquisition process: the EEG data were down-sampled to 250Hz and sent out every 40 ms, i.e., each DW update contained 10 samples ($L'=10$).

\subsection{Hyper-parameters}

There were two types of hyper-parameters: model training related, and FRDW related.

Model training related hyper-parameters included the trial length, FR ratio, maximum number of epochs, learning rate, and batch size. The data length was selected from $\{100, 125, 150, 200, 250, 500, 750\}$ according to the ITR on the validation set. The maximum number of training epochs 100, learning rate 0.001, and batch size 64 were used. To cope with randomness, in each experiment, EEGNet and CSP+Transformer were repeated 11 times, and the average performance was reported. CSP+SVM was run only once, as it was much more stable.

The values of the FRDW related hyper-parameters are shown in Table~\ref{tab:hyper-parameters}.

\begin{table}[h]\centering \setlength\tabcolsep{5mm}
\caption{FRDW related hyper-parameters.} \label{tab:hyper-parameters}
\begin{tabular}{@{}ccccc@{}}\toprule
\multirow{2}{*}{Setting} & \multicolumn{2}{c}{Binary classification} & \multicolumn{2}{c}{4-class classification} \\ \cmidrule(l){2-5}
                                    & $\underline{L}$          & $\tau$         & $\underline{L}$          & $\tau$          \\ \midrule
Within-subject                       & 60                    & 0.7               & 60                    & 0.6                \\
Cross-subject                       & 60                    & 0.6               & 50                    & 0.4                \\ \bottomrule
\end{tabular}
\end{table}

\section{Results}

This section presents the experimental results to demonstrate the effectiveness of FRDW.

\subsection{Online MI Classification}

Tables~\ref{tab:within on MI1}-\ref{tab:within on MI3} show the ITRs and accuracies (ACCs) in within-subject classification on the three datasets, respectively. Table~\ref{tab:within on MI1} also includes the detailed results on each individual subject. Tables~\ref{tab:cross on MI1}-\ref{tab:cross on MI3} show the ITRs and ACCs in cross-subject classification on three datasets, respectively. The best ITRs for each training data augmentation approach are marked in bold. The values in parentheses are the improvements of FRDW over FW.

\begin{table*}[h] \centering \setlength\tabcolsep{4pt}
\caption{ITRs and accuracies in within-subject classification on MI1.} \label{tab:within on MI1}
\begin{tabular}{cccccccccccc|cc} \toprule
\multirow{2}[2]{*}{Model}                                                      & \multirow{2}[2]{*}{\begin{tabular}[c]{@{}c@{}}Training Data \\ Augmentation\end{tabular}} & \multirow{2}[2]{*}{Test Strategy} & \multicolumn{9}{c}{Subject}                                                                                                                              & \multicolumn{2}{|c}{Avg.}              \\ \cmidrule(l){4-14}
&  &                                & 1               & 2             & 3               & 4              & 5             & 6              & 7               & 8               & 9       & \multicolumn{1}{c}{ITR}             & \multicolumn{1}{c}{ACC}                         \\ \midrule
\multirow{6}[6]{*}{EEGNet} & \multirow{2}[2]{*}{none} & FW                             & 76.42           & 5.43          & 92.72           & 36.27          & 6.57          & 12.67          & 67.68           & 77.72           & 62.93        & 48.71      & 0.54                       \\
&  & FRDW                          & 83.95           & 5.4           & 107.46          & 37.01          & \textbf{6.63} & 12.79          & 66.44           & 90.37           & 68.19     & \textbf{53.14 ($\uparrow$9.1\%)}      & 0.53 ($\downarrow$0.01)  \\ \cmidrule(l){2-14}
& \multirow{2}[2]{*}{overlap}  & FW                             & 95.61           & \textbf{9.62} & 92.72           & 39.91          & 0.01          & 13.21          & 83.01           & 89.89           & 66.48     & 54.49     & 0.55                           \\
&   & FRDW                          & 114.3           & 8.3           & 107.64          & \textbf{41.38} & 0.03          & 13.48          & 86.21           & 89.9            & \textbf{72.78}  & \textbf{59.33 ($\uparrow$8.9\%)}
& 0.53 ($\downarrow$0.02)      \\ \cmidrule(l){2-14}
& \multirow{2}[2]{*}{FR}  & FW                             & 101.53          & 8.24          & 95.61           & 38.98          & 0.25          & 14.34          & 85.72           & 94.16           & 67.68    & 56.28      & 0.56                           \\
&   & FRDW                          & \textbf{133.37} & 7.34          & \textbf{119.85} & 40.07          & 0.25          & \textbf{14.63} & \textbf{97.19}  & \textbf{115.37} & 70.5            & \textbf{66.51 ($\uparrow$18.2\%)}
& 0.55 ($\downarrow$0.01)      \\ \midrule \midrule
\multirow{6}[6]{*}{\begin{tabular}[c]{@{}c@{}}CSP+\\ Transformer\end{tabular}} & \multirow{2}[2]{*}{none}      & FW                             & 74.18           & 14.88         & 77.66           & 23.6           & \textbf{1.91} & 12.4           & 59.09           & 73.04           & 63.22          & 44.44  & 0.56                          \\
&       & FRDW                          & 101.19          & 6.64          & 99.46           & 31.09          & 1.07          & 13.93          & 106.79          & 77.22           & \textbf{77.19}& \textbf{57.18 ($\uparrow$28.7\%)}
& 0.53 ($\downarrow$0.03)       \\ \cmidrule(l){2-14}
& \multirow{2}[2]{*}{overlap}  & FW                             & 73.04           & \textbf{12.4} & 65.33           & 24.9           & 0.09          & 13.37          & 64.27           & 78.83           & 64.27    & 44.06      & 0.55                           \\
&   & FRDW                          & 108.9           & 7.34          & 90.84           & 33.13          & 0.05          & 13.6           & 107.64          & 76.32           & 73.04     & \textbf{56.76 ($\uparrow$28.8\%)}     & 0.52 ($\downarrow$0.03)  \\ \cmidrule(l){2-14}
& \multirow{2}[2]{*}{FR}   & FW                             & 73.04           & 6.6           & 78.83           & 25.57          & 0.35          & \textbf{18.14} & 73.04           & 76.49           & 64.27    & 46.26      & 0.56                           \\
&   & FRDW                          & \textbf{123.13} & 7.63          & \textbf{110.68} & \textbf{34.19} & 0.79          & 16.04          & \textbf{111.67} & \textbf{80.6}   & 72.08    & \textbf{61.87 ($\uparrow$33.7\%)}      & 0.55 ($\downarrow$0.01)  \\ \midrule \midrule
\multirow{6}[6]{*}{CSP+SVM}    & \multirow{2}[2]{*}{none}   & FW                             & 56.14           & \textbf{1.71} & 52.90           & 37.16          & 0.06          & 4.39           & 32.79           & 35.40           & 52.90           & 30.38     & 0.47                          \\
&    & FRDW                          & 74.38           & 1.42          & 65.44           & \textbf{37.54} & 0.06          & 4.39           & \textbf{44.79}  & 34.3            & 55.09     & \textbf{35.27 ($\uparrow$16.1\%)}     & 0.46 ($\downarrow$0.01)  \\ \cmidrule(l){2-14}
& \multirow{2}[2]{*}{overlap}     & FW                             & 64.11           & 0.97          & 52.90           & 34.52          & \textbf{1.31} & 4.07           & 33.65           & 33.65           & 50.77      & 30.66    & 0.47                           \\
&   & FRDW                          & \textbf{76.45}  & 0.74          & 66.31           & 34.86          & \textbf{1.31} & 4.11           & 48.9            & 34.7            & 52.34      & \textbf{35.53 ($\uparrow$15.9\%)}    & 0.46 ($\downarrow$0.01)  \\ \cmidrule(l){2-14}
& \multirow{2}[2]{*}{FR}  & FW                             & 61.76           & 0.97          & 52.90           & 31.97          & 0.97          & \textbf{4.73}  & 29.51           & 33.65           & 53.95      & 30.04    & 0.47                           \\
&  & FRDW                          & 67.8            & 1.22          & \textbf{67.8}   & 37.16          & 0.03          & 3.76           & 39.46           & \textbf{34.98}  & \textbf{55.63}
& \textbf{34.20 ($\uparrow$13.8\%)}      & 0.46 ($\downarrow$0.01)  \\ \bottomrule
\end{tabular}
\end{table*}

\begin{table}[h] \centering \setlength\tabcolsep{3pt}
\caption{Average ITRs and accuracies in within-subject classification on MI2.} \label{tab:within on MI2}
\begin{tabular}{cc|cc|cc|cc} \toprule
\multirow{3}{*}{Model}           & \multirow{3}{*}{} & \multicolumn{6}{c}{Training Data Augmentation}     \\ \cmidrule(l){3-8}
&                       & \multicolumn{2}{c}{none}       & \multicolumn{2}{|c|}{overlap}    & \multicolumn{2}{c}{FR}         \\ \cmidrule(l){3-8}
&  & FW    & FRDW                  & FW    & FRDW   & FW    & FRDW                  \\ \midrule
\multirow{4}[2]{*}{EEGNet}          & \multirow{2}{*}{ITR}                   & 21.26 & \textbf{25.91} & 25.09& \textbf{30.11} & 24.70 & \textbf{29.76} \\
 &                   &  & \textbf{($\uparrow$21.9\%)} &  & \textbf{($\uparrow$20.0\%)} & & \textbf{($\uparrow$20.5\%)} \\ \cmidrule(l){2-8}
 &\multirow{2}{*}{ACC}      & 0.66  & 0.66     & 0.70  & 0.68    & 0.69  & 0.68 \\
 &      &   & (--)  &  & ($\downarrow$0.02)   &  & ($\downarrow$0.01)             \\ \midrule
\multirow{4}[2]{*}{\begin{tabular}[c]{@{}c@{}}CSP+\\ Transformer\end{tabular}} &
\multirow{2}{*}{ITR}                   & 24.00 & \textbf{34.04} & 19.32 & \textbf{32.49} & 24.64 & \textbf{33.73} \\
& & & \textbf{($\uparrow$41.8\%)} & & \textbf{($\uparrow$68.2\%)} & & \textbf{($\uparrow$36.9\%)} \\ \cmidrule(l){2-8}
& \multirow{2}{*}{ACC}  & 0.70  & 0.69         & 0.69  & 0.69    & 0.71  & 0.69            \\
& &  & ($\downarrow$0.01)      &  & (--)             &  & ($\downarrow$0.02)             \\ \midrule
\multirow{4}[2]{*}{CSP+SVM}         & \multirow{2}{*}{ITR}  & 25.36 & \textbf{34.97} & 25.09 & \textbf{32.45} & 24.32 & \textbf{33.92} \\ \
&  &  & \textbf{($\uparrow$37.9\%)} & & \textbf{($\uparrow$29.3\%)} & & \textbf{($\uparrow$39.5\%)} \\ \cmidrule(l){2-8}
& \multirow{2}{*}{ACC}   & 0.70  & 0.69   & 0.69  & 0.68   & 0.69  & 0.69 \\
&  &  & ($\downarrow$0.01)     &  & ($\downarrow$0.01)   & & (--)          \\\bottomrule
\end{tabular}
\end{table}

\begin{table}[h] \centering \setlength\tabcolsep{3pt}
\caption{Average ITRs and accuracies in within-subject classification on MI3.} \label{tab:within on MI3}
\begin{tabular}{cc|cc|cc|cc}\toprule
\multirow{3}{*}{Model}          & \multirow{3}{*}{} & \multicolumn{6}{c}{Training Data Augmentation}                                                   \\ \cmidrule(l){3-8}
 &                       & \multicolumn{2}{c}{none}       & \multicolumn{2}{|c|}{overlap}    & \multicolumn{2}{c}{FR}         \\ \cmidrule(l){3-8}
  &                       & FW    & FRDW                  & FW    & FRDW                  & FW    & FRDW                  \\ \midrule
\multirow{4}[2]{*}{EEGNet}   & \multirow{2}{*}{ITR}                   & 21.19 & \textbf{34.10} & 23.20 & \textbf{34.05} & 22.82 & \textbf{33.81} \\
&   & & \textbf{($\uparrow$60.9\%)} &  & \textbf{($\uparrow$46.8\%)} & & \textbf{($\uparrow$48.2\%)} \\ \cmidrule(l){2-8}
& \multirow{2}{*}{ACC}  & 0.72  & 0.70   & 0.73  & 0.70  & 0.73  & 0.70          \\
& & & ($\downarrow$0.02)  &  & ($\downarrow$0.03)   &  &  ($\downarrow$0.03)       \\ \midrule
\multirow{4}[2]{*}{\begin{tabular}[c]{@{}c@{}}CSP+\\ Transformer\end{tabular}} & \multirow{2}{*}{ITR}                   & 26.10 & \textbf{32.11} & 27.48 & \textbf{30.76} & 25.86 & \textbf{32.69} \\
&  &  & \textbf{($\uparrow$23.0\%)} & & \textbf{($\uparrow$11.9\%)} & & \textbf{($\uparrow$26.4\%)} \\\cmidrule(l){2-8}
& \multirow{2}{*}{ACC}  & 0.72  & 0.71   & 0.73  & 0.68  & 0.72  & 0.68      \\
&   &  & ($\downarrow$0.01)   &  & ($\downarrow$0.05)  & & ($\downarrow$0.04)   \\ \midrule
\multirow{4}[2]{*}{CSP+SVM}       & \multirow{2}{*}{ITR}                   & 19.56 & \textbf{28.66} & 19.90 & \textbf{28.18} & 19.72 & \textbf{25.16} \\
&  &  & \textbf{($\uparrow$46.5\%)} & & \textbf{($\uparrow$41.6\%)} & & \textbf{($\uparrow$27.6\%)} \\ \cmidrule(l){2-8}
& \multirow{2}{*}{ACC}    & 0.71  & 0.68   & 0.71  & 0.68    & 0.70  & 0.68         \\
& & & ($\downarrow$0.03)  & & ($\downarrow$0.03)   &  & ($\downarrow$0.02)       \\ \bottomrule
\end{tabular}
\end{table}

\begin{table*}[h] \centering \setlength\tabcolsep{1pt}
\caption{Average ITRs and accuracies in cross-subject classification on MI1.} \label{tab:cross on MI1}
\begin{tabular}{cc|cccc|cccc|cccc} \toprule
\multirow{4}{*}{Model}           & \multirow{4}{*}{} & \multicolumn{12}{c}{Training Data Augmentation}                                                                                                                                                           \\ \cmidrule(l){3-14}
                                 &                       & \multicolumn{4}{|c|}{none}                                          & \multicolumn{4}{|c|}{overlap}                                       & \multicolumn{4}{c}{FR}                                            \\ \cmidrule(l){3-14}
                                 &                       & FW    & FRDW                   & EA+FW & EA+FRDW               & FW    & FRDW                   & EA+FW & EA+FRDW               & FW    & FRDW                   & EA+FW & EA+FRDW               \\ \midrule
\multirow{2}[2]{*}{EEGNet}          & ITR                   & 13.69 & 14.26 ($\uparrow$4.2\%)  & 19.40 & \textbf{27.83 ($\uparrow$43.5\%)} & 14.30 & 16.84 ($\uparrow$17.8\%) & 19.78 & \textbf{26.97 ($\uparrow$36.3\%)} & 11.84 & 14.01 ($\uparrow$18.3\%) & 18.22 & \textbf{26.37 ($\uparrow$44.7\%)} \\ \cmidrule(l){2-14}
                                 & ACC                   & 0.37  & 0.36 ($\downarrow$0.01)             & 0.41  & 0.40 ($\downarrow$0.01)             & 0.37  & 0.36 ($\downarrow$0.01)             & 0.42  & 0.40 ($\downarrow$0.02)             & 0.36  & 0.35 ($\downarrow$0.01)             & 0.41  & 0.39 ($\downarrow$0.01)             \\ \midrule
\multirow{2}[2]{*}{\begin{tabular}[c]{@{}c@{}}CSP+\\ Transformer\end{tabular}} & ITR                   & 12.55 & 16.41 ($\uparrow$30.8\%) & 16.67 & \textbf{25.00 ($\uparrow$50.0\%)} & 11.40 & 10.47 ($\downarrow$8.2\%)  & 15.70 & \textbf{23.42 ($\uparrow$49.2\%)} & 10.70 & 13.04 ($\uparrow$21.9\%) & 17.40 & \textbf{27.47 ($\uparrow$57.9\%)} \\ \cmidrule(l){2-14}
                                 & ACC                   & 0.38  & 0.36 ($\downarrow$0.02)             & 0.41  & 0.39 ($\downarrow$0.02)             & 0.37  & 0.33 ($\downarrow$0.04)             & 0.41  & 0.39 ($\downarrow$0.02)             & 0.37  & 0.35 ($\downarrow$0.02)             & 0.43  & 0.41 ($\downarrow$0.01)             \\ \midrule
\multirow{2}[2]{*}{CSP+SVM}         & ITR                   & 13.19 & 15.94 ($\uparrow$20.8\%) & 17.14 & \textbf{24.47 ($\uparrow$42.8\%)} & 12.90 & 14.84 ($\uparrow$15.0\%) & 17.04 & \textbf{24.52 ($\uparrow$43.9\%)} & 12.83 & 14.78 ($\uparrow$15.2\%) & 17.13 & \textbf{22.56 ($\uparrow$31.7\%)} \\ \cmidrule(l){2-14}
                                 & ACC                   & 0.38  & 0.37 ($\downarrow$0.01)            & 0.40  & 0.40 (-)            & 0.38  & 0.37 ($\downarrow$0.01)            & 0.40  & 0.40 (-)            & 0.38  & 0.37 (-)            & 0.40  & 0.40 (-)            \\ \bottomrule
\end{tabular}
\end{table*}

\begin{table*}[h] \centering \setlength\tabcolsep{1pt}
	\caption{Average ITRs and accuracies in cross-subject classification on MI2.} \label{tab:cross on MI2}
	\begin{tabular}{cc|cccc|cccc|cccc}\toprule
		\multirow{4}{*}{Model}           & \multirow{4}{*}{} & \multicolumn{12}{c}{Training Data Augmentation}                                                                                                                                                     \\ \cmidrule(l){3-14}
		&                       & \multicolumn{4}{c}{none}                                        & \multicolumn{4}{|c|}{overlap}                                     & \multicolumn{4}{c}{FR}                                          \\ \cmidrule(l){3-14}
		&                       & FW    & FRDW                  & EA+FW & EA+FRDW              & FW    & FRDW                  & EA+FW & EA+FRDW              & FW    & FRDW                  & EA+FW & EA+FRDW              \\ \midrule
		
		\multirow{2}[2]{*}{EEGNet}             & ITR   & 12.05 & \textbf{21.41 ($\uparrow$77.9\%)} & 8.08  & 12.41 ($\uparrow$53.4\%)          & 12.07 & \textbf{19.68 ($\uparrow$63.1\%)} & 12.27 & 15.14 ($\uparrow$23.4\%) & 13.38 & \textbf{17.25 ($\uparrow$28.9\%)} & 12.10 & 13.47 ($\uparrow$11.3\%)           \\ \cmidrule(l){2-14}
		& ACC   & 0.61  & 0.63 ($\uparrow$0.02)           & 0.61  & 0.61 (-)              & 0.63  & 0.63 (-)               & 0.64  & 0.63 ($\downarrow$0.01)  & 0.63  & 0.63 (-)               & 0.63  & 0.62 ($\downarrow$0.01)            \\ \midrule
		
		\multirow{2}[2]{*}{\begin{tabular}[c]{@{}c@{}}CSP+\\ Transformer\end{tabular}}   & ITR   & 11.69 & \textbf{21.55 ($\uparrow$84.4\%)} & 9.94  & 13.19 ($\uparrow$32.7\%)          & 11.59 & \textbf{18.38 ($\uparrow$58.6\%)} & 9.74  & 13.95 ($\uparrow$43.2\%) & 11.37 & \textbf{18.38 ($\uparrow$61.7\%)} & 11.22 & 17.68 ($\uparrow$57.6\%)          \\ \cmidrule(l){2-14}
		& ACC   & 0.62  & 0.61 ($\downarrow$0.01)           & 0.63  & 0.61 ($\downarrow$0.02)          & 0.60  & 0.61 ($\uparrow$0.01)           & 0.64  & 0.62 ($\downarrow$0.02)  & 0.61  & 0.61 ((-)              & 0.64  & 0.64 (-)               \\ \midrule
		
		\multirow{2}[2]{*}{CSP+SVM}            & ITR   & 10.05 & 12.75 ($\uparrow$26.9\%)          & 9.76  & \textbf{12.79 ($\uparrow$31.1\%)} & 9.79  & \textbf{13.64 ($\uparrow$39.3\%)} & 7.66  & 10.23 ($\uparrow$33.6\%) & 9.06  & 11.57 ($\uparrow$27.7\%)          & 9.94  & \textbf{13.51 ($\uparrow$35.9\%)} \\ \cmidrule(l){2-14}
		& ACC   & 0.60  & 0.59 ($\downarrow$0.01)           & 0.63  & 0.62 ($\downarrow$0.01)          & 0.59  & 0.59 (-)               & 0.61  & 0.61 (-)      & 0.60  & 0.59 ($\downarrow$0.01)           & 0.63  & 0.63 (-)               \\ \bottomrule
	\end{tabular}
\end{table*}

\begin{table*}[h] \centering \setlength\tabcolsep{1pt}
\caption{Average ITRs and accuracies in cross-subject classification on MI3.} \label{tab:cross on MI3}
\begin{tabular}{cc|cccc|cccc|cccc}\toprule
\multirow{4}{*}{Model}           & \multirow{4}{*}{} & \multicolumn{12}{c}{Training Data Augmentation}                                                                                                                                                     \\ \cmidrule(l){3-14}
&                       & \multicolumn{4}{c}{none}                                        & \multicolumn{4}{|c|}{overlap}                                     & \multicolumn{4}{c}{FR}                                          \\ \cmidrule(l){3-14}
&                       & FW    & FRDW                  & EA+FW & EA+FRDW              & FW    & FRDW                  & EA+FW & EA+FRDW              & FW    & FRDW                  & EA+FW & EA+FRDW              \\ \midrule

\multirow{2}[2]{*}{EEGNet}          & ITR                   & 20.51 & 33.63 ($\uparrow$64.0\%) & 21.12 & \textbf{34.67 ($\uparrow$64.2\%)} & 21.44 & 31.51 ($\uparrow$47.0\%) & 21.93 & \textbf{32.04 ($\uparrow$46.1\%)} & 21.89 & 34.32 ($\uparrow$56.8\%) & 22.31 & \textbf{35.96 ($\uparrow$61.2\%)} \\ \cmidrule(l){2-14}
                                 & ACC                   & 0.72  & 0.69 ($\downarrow$0.03)             & 0.72  & 0.7 ($\downarrow$0.02)              & 0.73  & 0.68 ($\downarrow$0.05)             & 0.72  & 0.69 ($\downarrow$0.04)             & 0.73  & 0.69 ($\downarrow$0.04)             & 0.72  & 0.70 ($\downarrow$0.02)             \\ \midrule
\multirow{2}[2]{*}{\begin{tabular}[c]{@{}c@{}}CSP+\\ Transformer\end{tabular}} & ITR                   & 19.37 & 25.53 ($\uparrow$31.8\%) & 20.73 & \textbf{31.23 ($\uparrow$50.7\%)} & 19.60 & 26.73 ($\uparrow$36.4\%) & 20.91 & \textbf{33.45 ($\uparrow$60.0\%)} & 18.90 & 26.48 ($\uparrow$40.1\%) & 19.44 & \textbf{29.09 ($\uparrow$49.6\%)} \\ \cmidrule(l){2-14}
                                 & ACC                   & 0.69  & 0.67 ($\downarrow$0.02)             & 0.70  & 0.69 ($\downarrow$0.01)             & 0.69  & 0.67 ($\downarrow$0.02)             & 0.70  & 0.69 ($\downarrow$0.01)             & 0.69  & 0.67 ($\downarrow$0.02)             & 0.69  & 0.69 ($\downarrow$0.01)             \\ \midrule
\multirow{2}[2]{*}{CSP+SVM}         & ITR                   & 17.07 & \textbf{33.77 ($\uparrow$97.8\%)} & 18.05 & 31.89 ($\uparrow$76.7\%) & 17.22 & \textbf{32.97 ($\uparrow$91.5\%)} & 17.87 & 31.37 ($\uparrow$75.5\%) & 17.24 & \textbf{32.82 ($\uparrow$90.4\%)} & 17.92 & 32.11 ($\uparrow$79.2\%) \\ \cmidrule(l){2-14}
                                 & ACC                   & 0.69  & 0.68 ($\downarrow$0.01)            & 0.70  & 0.68 ($\downarrow$0.02)             & 0.69  & 0.68 ($\downarrow$0.02)            & 0.70  & 0.68 ($\downarrow$0.02)             & 0.69  & 0.67 ($\downarrow$0.02)            & 0.70  & 0.69 ($\downarrow$0.01)             \\ \bottomrule
\end{tabular}
\end{table*}

Tables~\ref{tab:within on MI1}-\ref{tab:cross on MI3} show that:
\begin{enumerate}
   \item Compared with FW, FRDW improved ITR with little or no loss of accuracy, regardless of the dataset, classifier, and data augmentation approach.
   \item FR can also be used for data augmentation in training, achieving comparable or better results than others.
    \item In cross-subject classification (Tables~\ref{tab:cross on MI1}-\ref{tab:cross on MI3}), combining EA with FRDW can further improve the ITR in most cases.
\end{enumerate}

To examine if the differences between FW and our proposed FRDW, and between FRDW without and with EA, were statistically significant, we performed paired-sample $t$-tests on the ITRs in Tables~\ref{tab:within on MI1}-\ref{tab:cross on MI3}. The null hypothesis was that the difference between the paired samples has zero mean, and it was rejected if $p \leq 0.05$.

The within- and cross-subject paired-sample $t$-test results between FW and FRDW are shown in Tables~\ref{tab:intra_ttest} and \ref{tab:cross_ttest}, respectively. The results for FRDW without and with EA are shown in Table~\ref{tab:ea_TTEST}. The statistically significant ones are marked in bold.

Table~\ref{tab:intra_ttest} shows that most $p$-values were smaller than or close to 0.05 when FR was used for train data augmentation, indicating statistically significant ITR improvement of FRDW over FW in within-subject MI classification. Table~\ref{tab:cross_ttest} shows that most $p$-values were smaller than or close to 0.05 when FR and EA were used together, indicating statistically significant ITR improvement of FRDW over FW in cross-subject MI classification.

%%%%%%%%%%%%within-subject ttest
\begin{table*}[h]   \centering
  \caption{Paired-sample $t$-test results on the test ITRs in Tables~\ref{tab:within on MI1}-\ref{tab:within on MI3} between FW and FRDW.}
    \begin{tabular}{c|ccc|ccc|ccc}    \toprule
    \multirow{3}[6]{*}{Model} & \multicolumn{9}{c}{Training Data Augmentation} \\
\cmidrule{2-10}        & \multicolumn{3}{c|}{none} & \multicolumn{3}{c|}{overlap} & \multicolumn{3}{c}{FR} \\
\cmidrule{2-10}        & MI1 & MI2 & MI3 & MI1 & MI2 & MI3 & MI1 & MI2 & MI3 \\
    \midrule
    EEGNet & 0.058  & \textbf{0.007} & \textbf{0.039} & 0.079  & 0.232  & 0.074  & \textbf{0.040} & \textbf{0.034} & 0.075  \\
    CSP+Transformer & 0.057  & 0.162  & 0.067  & 0.065  & \textbf{0.017} & 0.263  & \textbf{0.041} & \textbf{0.049} & 0.073  \\
    CSP+SVM & 0.079  & 0.072  & 0.098  & 0.060  & 0.110  & 0.106  & \textbf{0.050} & \textbf{0.049} & 0.232  \\     \bottomrule
    \end{tabular}   \label{tab:intra_ttest}%
\end{table*}%

%%%%%%%%%%%%cross-subject ttest
\begin{table*}[h]   \centering \setlength\tabcolsep{3pt}
  \caption{Paired-sample $t$-test results on the test ITRs in Tables~\ref{tab:cross on MI1} and~\ref{tab:cross on MI3} between FW and FRDW.}
    \begin{tabular}{c|ccc|ccc|ccc|ccc|ccc|ccc}     \toprule
    \multirow{3}[6]{*}{Model} & \multicolumn{18}{c}{Training Data Augmentation} \\
\cmidrule{2-19}        & \multicolumn{3}{c|}{none w/o EA} & \multicolumn{3}{c|}{none w/ EA} & \multicolumn{3}{c|}{ overlap w/o EA} & \multicolumn{3}{c|}{overlap w/ EA} & \multicolumn{3}{c|}{FR w/o EA} & \multicolumn{3}{c}{FR w/ EA} \\
\cmidrule{2-19} & MI1 & MI2 & MI3 & MI1 & MI2 & MI3 & MI1 & MI2 & MI3 & MI1 & MI2 & MI3 & MI1 & MI2 & MI3 & MI1 & MI2 & MI3 \\   \midrule
    EEGNet & 0.641 & 0.074 & \textbf{0.007}    & 0.077 & 0.094 & 0.053    & 0.176 & 0.149 & 0.232    & \textbf{0.013} & 0.170 & 0.053    & 0.085 & \textbf{0.035} & \textbf{0.034}  & 0.079  &0.099 & \textbf{0.041} \\
    CSP+Transformer & 0.134 & 0.234 & 0.162    & 0.080 & 0.168 & 0.061    & 0.527 & 0.185  & \textbf{0.017}   & 0.059 & \textbf{0.023}  & \textbf{0.043}   & 0.322 & 0.089 & \textbf{0.049}  & \textbf{0.014} & 0.087 & 0.377  \\
    CSP+SVM & 0.241 &0.278 & 0.072    & \textbf{0.028} &\textbf{0.037} & 0.066   & 0.192 &0.291 & 0.110    & \textbf{0.020} &\textbf{0.020} & 0.063   & 0.182 &0.281 & \textbf{0.049}  & \textbf{0.034} &\textbf{0.029} & 0.060  \\     \bottomrule
    \end{tabular}   \label{tab:cross_ttest}%
\end{table*}%

%%%%%%%%%%%%EA ttest
\begin{table}[h]  \centering \setlength\tabcolsep{10pt}
  \caption{Paired-sample $t$-test results on the test ITRs in Tables~\ref{tab:cross on MI1} and \ref{tab:cross on MI3} for FRDW without EA and with EA.}
    \begin{tabular}{ccccc}    \toprule
    \multirow{2}[4]{*}{Model}  & \multirow{2}[4]{*}{Dataset}&\multicolumn{3}{c}{Training Data Augmentation} \\
\cmidrule{3-5}        &  & none & overlap & FR \\   \midrule
    \multirow{3}[2]{*}{EEGNet} & MI1  & 0.089  & \textbf{0.040} & 0.075  \\
    & MI2  & 0.199 & 0.381  & 0.324  \\
        & MI3  & 0.299  & 0.758  & 0.591  \\     \midrule
    \multirow{3}[2]{*}{CSP+Transformer} & MI1  & 0.919  & \textbf{0.021} & \textbf{0.007} \\
       & MI2  &0.399  &0.489  &0.891   \\
        & MI3  & \textbf{0.032} & 0.069  & 0.377  \\     \midrule
    \multirow{3}[2]{*}{CSP+SVM} & MI1  & \textbf{0.029} & \textbf{0.039} & 0.075  \\
    & MI2  &0.996  & 0.681 & 0.751  \\
        & MI3  & 0.519  & 0.636  & 0.832  \\     \bottomrule
    \end{tabular}   \label{tab:ea_TTEST}%
\end{table}%

\subsection{Sensitivity Analysis} \label{Sensitivity Analysis}

\begin{figure}[h]\centering
	\subfigure[]{\includegraphics[width=.48\linewidth,clip]{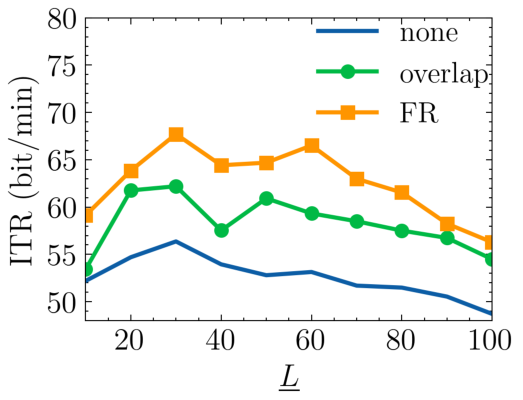}}
	\subfigure[]{\includegraphics[width=.48\linewidth,clip]{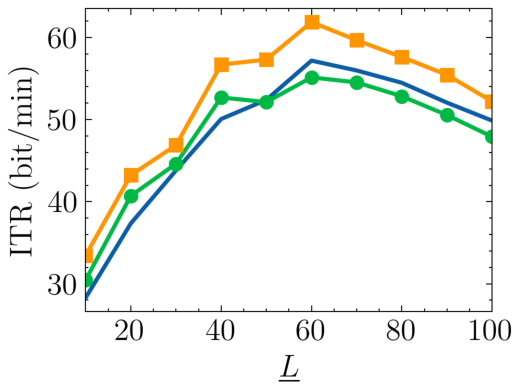}}
	\subfigure[]{\includegraphics[width=.48\linewidth,clip]{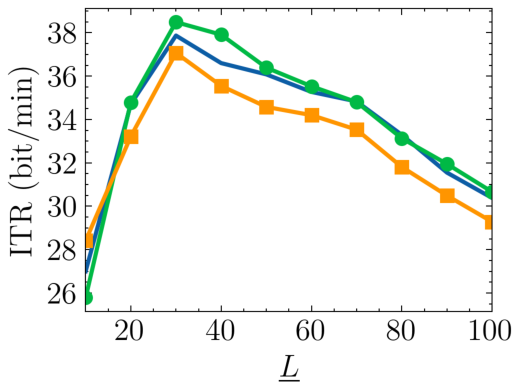}}
	\subfigure[]{\includegraphics[width=.48\linewidth,clip]{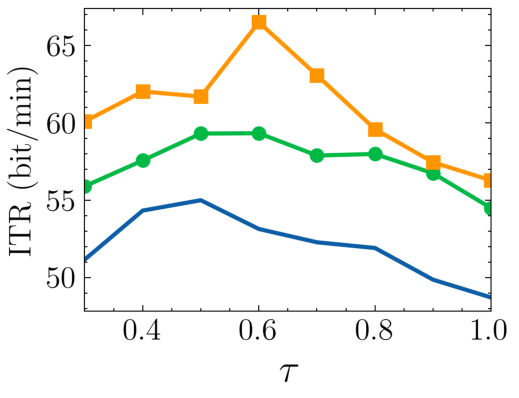}}
	\subfigure[]{\includegraphics[width=.48\linewidth,clip]{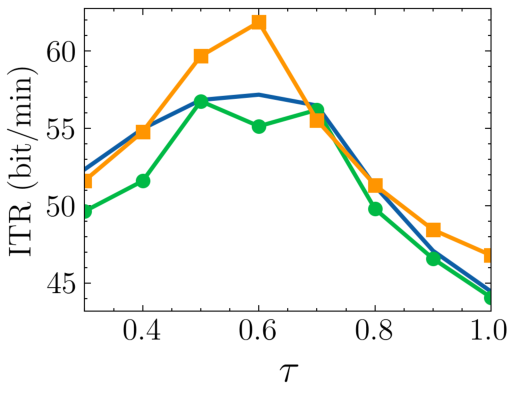}}
	\subfigure[]{\includegraphics[width=.48\linewidth,clip]{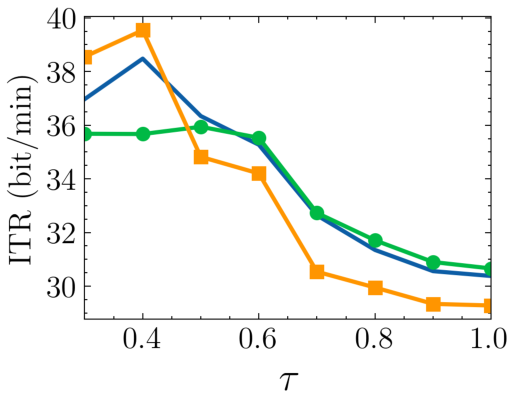}}
\caption{Sensitivity analysis on MI1 for within-subject classification. (a)-(c) ITR w.r.t. different $\underline{L}$ ($\tau=0.6$) on EEGNet, CSP+Transformer and CSP+SVM, respectively; (d)-(f) ITR w.r.t. different $\tau$ ($\underline{L}=60$) on EEGNet, CSP+Transformer and CSP+SVM, respectively.}     \label{fig:sensitive}
\end{figure}

Fig.~\ref{fig:sensitive} shows the sensitivity analysis results of the two hyper-parameters in FRDW, $\underline{L}$ and $\tau$, on MI1 for within-subject classification.

For a fixed confidence threshold $\tau$, when the minimum trial length $\underline{L}$ increased, generally the ITR first increased and then decreased. For a fixed minimum trial length $\underline{L}$, when the confidence threshold $\tau$ increased, the ITR also first increased and then decreased. Both results are intuitive.

Fig.~\ref{fig:sensitive} also shows that FR data augmentation almost always outperformed `none' and `overlap', except when CSP+SVM was used. However, since CSP+SVM generally performed worse than EEGNet and CSP+Transformer, EEGNet with FR or CSP+Transformer with FR is recommended.

\subsection{Ablation Study}

Transformer and SVM classifiers do not require the input EEG trial to have a fixed length, so they can also be used without data replication. However, this subsection shows that using FR in testing still improved their ITRs.

Tables~\ref{tab:AblationTransfomer}-\ref{tab:AblationSVM} show the test results without and with FR when Transformer and SVM were used as the classifier, respectively. The best ITRs are marked in bold. Using FR in testing always improved the ITRs for both classifiers.

%%%%%%%%%%%%%%Ablation CSP+Transfomer
\begin{table}[h] \centering
\caption{Average performance without and with FR in testing, when CSP+Transfomer was used on MI1.}
\label{tab:AblationTransfomer}
\begin{tabular}{l|cc|cc|cc} \toprule
\multirow{4}[4]{*}{Metric} & \multicolumn{6}{c}{Training Data Augmentation}                                           \\ \cmidrule(l){2-7}
& \multicolumn{2}{c}{none} & \multicolumn{2}{|c|}{overlap} & \multicolumn{2}{c}{FR}  \\ \cmidrule(l){2-7}
& w/o FR  & w/ FR          & w/o FR   & w/ FR            & w/o FR & w/ FR          \\ \midrule
ACC        & 0.53    & 0.53          & 0.52     & 0.52             & 0.55   & 0.55           \\ \midrule
ITR        & 55.16   & \textbf{57.18} & 54.96    & \textbf{56.76}   & 58.73  & \textbf{61.87} \\ \midrule
Time (s)                   & 0.39   & 0.38          & 0.36    &  0.36           & 0.40      &  0.39  \\ \bottomrule
\end{tabular}
\end{table}

%%%%%%%%%%%%%%Ablation CSP+SVM
\begin{table}[h] \centering
\caption{Average performance without and with FR in testing, when CSP+SVM was used on MI1.}
\label{tab:AblationSVM}
\begin{tabular}{l|cc|cc|cc} \toprule
\multirow{4}{*}{Metric} & \multicolumn{6}{c}{Training Data Augmentation}                                           \\ \cmidrule(l){2-7}
& \multicolumn{2}{c}{none} & \multicolumn{2}{|c|}{overlap} & \multicolumn{2}{c}{FR}  \\ \cmidrule(l){2-7}
& w/o FR  & w/ FR          & w/o FR            & w/ FR   & w/o FR & w/ FR          \\ \midrule
ACC        & 0.46    & 0.46           & 0.46              & 0.46    & 0.46   & 0.46           \\ \midrule
ITR        & 35.11   & \textbf{35.27} & 35.26    & \textbf{35.53}   & 33.94  & \textbf{34.20} \\ \midrule
Time (s)                 & 0.36   & 0.36          & 0.35             & 0.35   & 0.36  & 0.36          \\ \bottomrule
\end{tabular}
\end{table}

\subsection{Computational Cost}

The sampling rates of all three datasets were 250 Hz. We assumed that the BCI system brings in 10 sampling points ($L'=10$) each time, i.e., we update the test trial every 40 ms. FRDW is required to complete all computations within 40 ms for real-time operations.

Table~\ref{timecost} shows the mean and standard deviation of the FRDW computation time for each update on MI1. All experiments were implemented via Python 3.8 and PyTorch, and ran on a server with NVIDIA RTX 3090 GPU and Intel(R) Xeon(R) Gold 6226R 2.90GHz CPU. On average, FRDW always finished all computations within 40 ms, especially when EEGNet or CSP+SVM was used.

%%%%%%%%%%%%Computing time
\begin{table}[h]   \centering \setlength\tabcolsep{10pt}
  \caption{Computation time (ms) of FRDW.} \label{timecost}
    \begin{tabular}{cccc}     \toprule
    \multirow{2}[2]{*}{Model} & \multicolumn{1}{c}{\multirow{2}[2]{*}{\begin{tabular}[c]{@{}c@{}}Training Data \\ Augmentation\end{tabular}}} & \multicolumn{2}{c}{FRDW} \\
\cmidrule{3-4}       &    & Mean & Std \\     \midrule
    \multirow{3}[2]{*}{EEGNet} & none & 12.7  & 2.8  \\
       & overlap & 12.7  & 2.7  \\
       & FR & 12.5  & 2.6  \\     \midrule
    \multirow{3}[2]{*}{CSP+Transfomer} & none & 20.3  & 16.3  \\
       & overlap & 22.6  & 16.7  \\
       & FR & 20.1  & 17.9  \\     \midrule
    \multirow{3}[1]{*}{CSP+SVM} & none & 2.8  & 2.8  \\
       & overlap & 3.9  & 2.5  \\
       & FR & 4.3  & 2.9  \\     \bottomrule
    \end{tabular}%
\end{table}%

\section{Conclusions}

MI is a classical and popular paradigm in EEG-based BCIs. Online accurate and fast decoding is very important to its successful applications. This paper has proposed a simple yet effective FRDW algorithm for this purpose. Dynamic windows enable the classification based on a test EEG trial shorter than those used in training, improving the decision speed; front-end replication fills a short test EEG trial to the length used in training, improving the classification accuracy. Within-subject and cross-subject online MI classification experiments on three public datasets, with three different classifiers and three different data augmentation approaches, demonstrated that FRDW can significantly increase the information transfer rate in MI decoding. Additionally, FR can also be used in training data augmentation. FRDW helped win national champion of the China BCI Competition in 2022.

%\bibliographystyle{IEEEtran} \bibliography{cxrbib}


\end{document}